\documentclass[preprint,onecolumn,12pt,aps,prb,showpacs,showkeys]{revtex4-1}
\usepackage{graphicx}
\usepackage{color}
\usepackage{amsmath}
\usepackage{float}

\begin{document}

\preprint{BAQ/ 02 2017/ DFM}

\title{Superconductivity under pressure: application of the functional derivative}

\author{G.I. Gonz\'alez-Pedreros}

\author{R. Baquero}

\address{Departamento de F\'isica, Cinvestav.
Av. IPN 2508, GAM, 07360 Ciudad de M\'exico, M\'exico}.

\email{rbaquero@fis.cinvestav.mx}

\date{\today}

\begin{abstract} 

In this paper, we calculate the superconducting critical temperature as a function of pressure, $T_c(P)$, using a method based on the functional derivative of the critical temperature with the Eliashberg function, $\delta T_c/\delta \alpha^2F(\omega)$. The coulomb electron-electron repulsion parameter, $\mu^*(P)$, at each pressure is obtained in a consistent way by solving the linearized Migdal-Eliashberg equation. This method requires as the starting input only the knowledge of $T_c(P)$ at the starting pressure. It applies to superconductors for which the Migdal-Eliashberg equations hold. We study Al, a typical BCS weak coupling superconductor with a low $T_c$. Our results of $T_c(P)$ as a function of pressure for Al  show an excellent agreement with the calculations of Profeta $\textit {et al.}$ (Phys. Rev. Lett.{\bf 96}, 047003 (2006)) which agree well with experiment.	.

\end{abstract}



\maketitle

\section{Introduction}
To determine the superconducting critical temperature, $T_c$,  as a function of pressure we use the
 density functional theory (DFT) and the density functional perturbation theory~\cite{Baroni1,Baroni2,Giannozzi2} (DFPT) to get the electron and the phonon band structures and the  Eliashberg function $\alpha^2F(\omega)$ from first principles. We use the Quantum Espresso suite  codes \cite{democritos} for that purpose. This method applies to superconductors for which the Migdal-Eliashberg (ME) equations \cite{Eliashberg1,Eliashberg2} are valid to describe their superconducting properties as the electron-phonon ones. There is a set of parameters that influence each other when the ME equations are used, namely, the critical temperature, $T_c$, the electron-phonon interaction parameter,  $\lambda$, the Coulomb  electron-electron repulsion parameter, $\mu^*$, and the frequency at which the sum over the Matsubara frequencies is stopped, the so-called, cut-off frequency, $\omega_c$, which can actually be fixed numerically. We can take $\lambda$  from specific heat and  $T_c$ from resistivity experiments, for example. Then,  $\mu^*$ can be fitted to $T_c$ by solving the Linear Migdal-Eliashberg (LME) equation. In cases, where two of these parameters are unknown (usually $T_c$ and  $\mu^*$) a problem arises. To a certain extend, this is an unsolved problem. Oliveira et al. \cite{Oliveira} presented  a formulation of this problem that does not use the parameter  $\mu^*$. There are several suggestions in the literature on how to estimate this parameter.  From the solution of the LME equation, we can get  the coulomb electron-electron repulsion parameter, $\mu^*$, as long as we know  $T_c$, assuming that the  Eliashberg function is known and $\omega_c$ is fixed. There are other ways to estimate $\mu^*$. Morel and Anderson\cite{Morel} suggest the following analytic formula

\begin{equation}\label{and}
\mu^{*}=\frac{\mu}{1+\mu ln\left(\frac{E_{el}}{\omega_{ph}}\right)}
\end{equation}

where the dimensionless parameter $\mu=\langle V \rangle N(E_F)$ is the product of the averaged screened Coulomb interaction, V,  and the density of states at the Fermi energy, $N(E_F)$; $E_{el}$ and $\omega_{ph}$ are the electron and phonon energy scales, respectively.
Further, Bennemann and Garland~\cite{Bennemann}, Smith~\cite{Smith2} and Neve \textit{et all.}~\cite{Neve} give semi-empirical formulas to  estimate the behavior of the Coulomb pseudo-potential as a function of pressure, Liu \textit{et all.}~\cite{Liu} and Freericks \textit{et all.}~\cite{Freericks} calculate $\mu^*$ scaled to the maximum phonon frequency, meaning to replace $\omega_{ph}$ in Eq.(\ref{and}) by $\omega_{max}$, the maximum  phonon frequency.  Daams and Carbotte~\cite{Daams1} fit  $\mu^*$ solving the  LME equation using the experimental value of $T_c$. In a more recent work Bauer \textit{et all.}~\cite{Bauer} calculated corrections to $\mu^*$  based on the Hubbard-Holstein model. There is no consensus  concerning the proper way to estimate or to  calculate $\mu^*$ under pressure or even at ambient pressure. For example, for Nb at ambient pressure,  a set of different values for $\mu^*$ are reported : 0.117~\cite{Daams10}, 0.13~\cite{Neve}, 0.14~\cite{Lee}, 0.183~\cite{Butler}, 0.21~\cite{Savrasov} and 0.249~\cite{Ostanin} which differ considerably from each other. 

In this paper, we consider fcc Al. We start our calculation from the data at ambient pressure, say $P_i$, where $T_c(P_i)$, the crystal structure of the system and the lattice parameters at the first pressure are known. We first optimize the lattice parameters using the Quantum Espresso code~\cite{democritos}.  
So we start with lattice parameters that minimize the energy as a function of the volume. We then obtain $ \alpha^2F(\omega, P_i)$. $\mu^*(P_i)$ is fitted to $T_c(P_i)$ solving the  LME equation. We fix $\omega_c$=10 $\omega_{max}$, the maximum phonon frequency. 
 We solve, at $P_i$, the  LME equation using the Mc Master programs~\cite{Daams1, Daams10, Leavens, Daams2, Daams3, carbotte00}. We obtain, at $P_i$, $\mu^*(P_i)$ and then the functional derivative   $\delta T_c/\delta \alpha^2F(\omega,P_i)$ ~\cite{Bergmann}. 

We define a next pressure, say $P_{i+1}$ and obtain the Eliashberg function at this new pressure. The $T_c (P_{i+1})$ is obtained from the value of the functional derivative at $P_i$  and the difference in the Eliashberg functions at the two pressures considered  (see below for details). From the knowledge of $T_c(P_{i+1})$ we fit the value of $\mu^*(P_{i+1})$ by solving the LME equation which we then use to obtain $\delta T_c/\delta \alpha^2F(\omega,P_{i+1})$.   This procedure can be repeated to get $T_c$ at other pressures. One has to be careful with the magnitude of the interval at which we calculate the next pressure since the information carried through the functional derivative could become meaningless for too large pressure intervals.

The rest of the paper is organized as follows. In Section II,  we present the theory that supports our method. The method is described in detail in Section III.  In  Section IV, we report some technical details used in the calculation. In the next section V,  we present our results and compare them with other work, namely, with the known successful calculations of Profeta et al.~\cite{Profeta}, and with experiment ~\cite{Levy,Gubser,Sundqvist}. We present our conclusions in a final Section VI.

\section{The theory}

As we mentioned above, we solve the LME equation to fit ~$\mu^*(P)$ to the calculated value of $T_c(P)$.  On the imaginary axis, the LME equation  is

\begin{equation}\label{LEGE}
\rho\bar{\Delta}_n=\pi T \sum_{m}\left[(\lambda_{mn}-\mu^*)-\delta_{nm}\frac{|\tilde{\omega}_n|}{\pi T}\right]\bar{\Delta}_m,
\end{equation}

\begin{eqnarray}
\tilde{\omega}_n=\omega_n+\pi T \sum_{m}\lambda_{mn}sgn({\omega}_n),\\
\omega_n=(2n-1)\pi T,\\
\bar{\Delta}_n=\frac{\tilde{\Delta}_n}{\rho+|\tilde{\omega}_n|},
\end{eqnarray}

\begin{equation}
\lambda_{mn}=2\int_0^\infty\frac{d\omega \omega \alpha^2F(\omega)}{\omega^2+(\omega_n-\omega_m)^2}.
\end{equation}

where $T$ is the temperature, $\tilde{\Delta}_n$ is the gap function, $\omega_n$ is the Matsubara frequency, $\rho$ is the pair breaking parameter  and $n=0,\pm 1, \pm 2, ...$. In particular, $\lambda_{nn} \equiv \lambda$ is the electron-phonon coupling constant.

The numerical  solution of the LME, Eq. (\ref{LEGE}) requires  the summation over the Matsubara frequencies to be stopped at $\omega_c$ as we mentioned before. The error caused by this restriction can be compensated~\cite{Bergmann} by replacing the true Coulomb repulsion parameter $\mu$ by the pseudo-repulsion parameter  $\mu^*$ which we mentioned above and used in our calculations . Bergmann and Rainer~\cite{Bergmann} suggest a cut-off frequency ten times the maximum phonon frequency, $\omega_{max}$. Other authors consider that 3-7 could be enough~\cite{Daams10,Yansun}. The proper cut-off can be fixed numerically by studying the contribution of the last term in the summation. The  Eliashberg function is defined as follows 

\begin{eqnarray}
\nonumber
\alpha^2F(\omega)=\frac{1}{N(\epsilon_{F})}\sum_{mn}\sum_{\textbf{q}\nu}\delta(\omega-\omega_{\textbf{q}\nu})\sum_{\textbf{k}}|g_{\textbf{k}+\textbf{q},\textbf{k}}^{\textbf{q}\nu,mn}|^2\\
\times\delta(\epsilon_{\textbf{k}+\textbf{q},m}-\epsilon_{F})\delta(\epsilon_{\textbf{k},n}-\epsilon_{F}),
\end{eqnarray}

where $g_{\textbf{k}+\textbf{q},\textbf{k}}^{\textbf{q}\nu,mn}$ is the electron-phonon coupling matrix element, $\epsilon_{\textbf{k}+\textbf{q},m}$ and $\epsilon_{\textbf{k},n}$ are the energy of the quasi-particles in bands $m$ and $n$ with wave vectors $\textbf{k}+\textbf{q}$ and $\textbf{k}$, respectively.  $\omega_{\textbf{q}\nu}$ is the phonon energy with momentum  $\textbf{q}$ and branch  $\nu. $ $N(\epsilon_{F})$ is the electronic density of states at the  Fermi energy, $\epsilon_{F}$.

From the first order derivative of the self-consistent Kohn-Sham~\cite{Hohenberg, Kohn} (KS) potential, $V_{KS}$, with respect to the atomic displacements $\vec{u}_{s\textbf{R}}$ for the $s^{th}$ atom in the position $\textbf{R}$, the electron-phonon matrix element can be obtained as

\begin{equation}
g_{\textbf{k}+\textbf{q},\textbf{k}}^{\textbf{q}\nu,mn}=
\left(\frac{\hbar}{2\omega_{\textbf{q}\nu}}\right)^{1/2}
\left\langle\psi_{\textbf{k}+\textbf{q},m}|\Delta V_{KS}^{\textbf{q}\nu}|\psi_{\textbf{k},n}\right\rangle,
\end{equation}

where $\Delta V_{KS}^{\textbf{q}\nu}$ is the self-consistent first variation of the KS potential and $\psi_{\textbf{k},n}$ is the $n^{th}$ valence KS orbital of wave vector $\textbf{k}$.

The functional derivative of  $T_c$ with respect to ~$\alpha^2F( \omega)$, 
~$\delta T_c/\delta\alpha^2F(\omega)$, is central to this work.  With the algorithm of Bergmann and Rainer~\cite{Bergmann} and Leavens~\cite{Leavens} the functional derivative can be calculated. Several authors have worked this calculation from the solution of the LME equation~\cite{Daams1,Daams2,Daams3,Daams10}, as well as Baquero \textit{et all.}~\cite{carbotte00} and Yamsun \textit{et all.}~\cite{Yansun} as we mentioned before.

\begin{equation}
\frac{\delta T_c}{\delta \alpha^2F(\omega)}=-\frac{\delta \rho/\delta \alpha^2F(\omega)}{(\partial \rho/\partial T)_{T_c}}
\end{equation}

Ounce the functional derivative, ~$\delta T_c/\delta\alpha^2F( \omega)$, is known the change in $T_c$, $\Delta T_c$ , caused by a change in $ \alpha^2F(\omega)$ can be obtained directly  as we show next. 

The transition temperature of a superconductor depends on the effective interaction with the existing phonons in the system.
To have a high frequency phonon is not enough for a system to have a high-Tc  as it can be seen in Al where a 41 meV peak in the phonon spectrum is notorious. The functional derivative ${\delta T_c}/{\delta \alpha^2F(\omega)}$ shows how the different phonon frequencies participate in defining the $T_c$. As a function of the dimensionless variable $\hbar \omega/K_B T_c$ it presents a maximum at  about 7-8  which turns out to be universal for the conventional superconductors \citep{Daams10, Daams2} as Al where the electron-phonon interaction is known to be the mechanism. This defines the so called optimum frequency, $\omega_{opt}$. This is actually the most important phonon frequency as far as the magnitude of $T_c$ is concerned.  At any frequency, it shows how sensitive $T_c$ is to a change in $\alpha^2F(\omega)$ at this particular frequency. By applying pressure, we induce changes in the Eliashberg function. When $\alpha^2F(\omega)$ is changed by a certain amount the difference in the Eliashberg function, $\Delta\alpha^2F(\omega)$, together with the functional derivative allow to calculate the change in $T_c$, $\Delta T_c$, which is given by the formula~\citep{Bergmann, Baquero2}

\begin{equation}\label{diff}
\Delta T_{c, P_{i+1}, P_i}=\int^\infty_0\frac{\delta T_c}{\delta \alpha^2F(\omega)}\left[(\alpha^2F(\omega, P_{i+1})-\alpha^2F(\omega, P_{i})\right] d\omega.  
\end{equation}

There are several papers in the literature that deal with the functional derivative for different purposes. For example, Bergmann and Rainer~\cite{Bergmann} discuss how  $T_c$ is influenced by different parts of $\alpha^2F(\omega)$ and apply their findings to several crystalline and amorphous superconductors. Mitrovic~\cite{Mitrovic} considers it as a diagnostic tool to analyze the behavior of $T_c$ as a function of an external variable. Allen and Dynes~\cite{Allen3} study in detail the case of Pb, Baquero \textit{et all.}~\cite{Baquero} took several Eliashberg functions from experiment to study the changes in $T_c$ when $Nb_3Ge$ is taken off stoichiometry. Yansun \textit{et all.}~\cite{Yansun} investigated the superconducting properties of Li as a function of pressure at the interval of pressure where it undergoes three phase transitions. Mitrovic~\cite{Mitrovic} developed a general formalism to calculate the functional derivative of $T_c$ with respect to $\alpha^2F(\omega)$ for a superconductor with several bands with isotropic intra-band and inter-band interactions. 

To get $T_{c, P_{i+1}}$ at the next pressure we start from Eq.$\ref{diff}$ and use the next Eq.$\ref{tc+}$

\begin{equation}\label{tc+}
T_{c, P_{i+1}} = T_{c, P_i}+\Delta T_{c, P_{i+1}, P_i}
\end{equation}

which can in turn be used to fit $\mu^*(P_{i+1})$ using the LME equation. This procedure can be repeated at will. Our results are presented in several Tables below.

\section{Technical details} 

The electron and phonon (PHDOS) densities of states  for Al  have been calculated using the DFT and the DFPT with plane waves (PW) pseudo-potentials~\cite{Baroni1,Baroni2,Giannozzi2,Giannozzi}.  To calculate the density of states (DOS) a kinetic energy cut-off of 50 Ry was used.  Our calculations were performed using the  generalized gradient approximation (GGA) and the norm conserving pseudo-potential together with the plane wave self-consistent field (PWSCF)~\cite{Baronis}. For the electronic and vibrational calculations we used  a 32x32x32  and 16x16x16 Monkhorst-Park~\cite{Monkhorst} (MP) \textit{k} mesh, respectively. The PHDOS was obtained from individual phonons calculated on a 8x8x8 MP \textit{q} mesh using the tetrahedron method~\cite{MacDonald}. We used the Quantum Espresso code~\cite{democritos} for all these calculations.

\section{Results and discussion}

We now apply the method to the weak coupling superconductor Al. We have taken $T_c=$1.8K~\cite{Gubser} at ambient pressure which is our starting pressure. The Eliashberg function was obtained using the Espresso code. $\lambda$ was calculated directly from it, $\mu^*$ was fitted to $T_c$ using the LME equation. The functional derivative at this starting pressure was calculated using the Mc Master programs~\cite{Daams1, Daams10, Leavens, Daams2, Daams3, carbotte00}. 

\begin{figure}[H]
\begin{center}
\includegraphics[width=120mm]{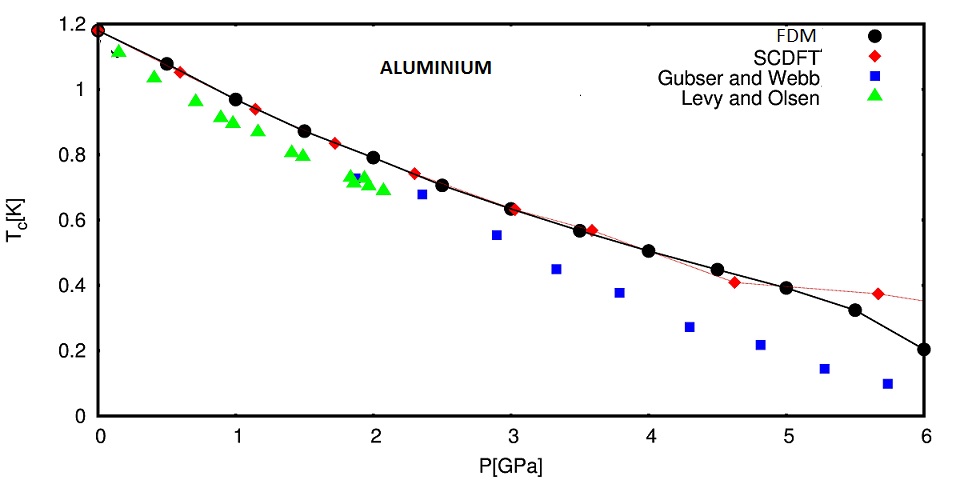}
\caption{\label{aluminio}( color on line) $T_c$[K] under pressure for Al: FDM is the present work, SCDFT is the linear interpolation of the data reported by Profeta \textit{et al.}~\cite{Profeta}.  Also the curves identified as Gusber and Webb~\cite{Gubser} and Levy and Olsen~\cite{Levy} are interpolations to their experimental data.}
\end{center}
\end{figure}

 From these starting data, we can obtain the variation of $T_c$ with pressure by applying the method just described. Other authors have worked in this problem. Namely, Dacorogna \textit{et all.}~\cite{Dacorogna} have calculated the $T_c$ as a function of pressure. They calculated self-consistently the phonon frequencies and the electron-phonon coupling. $\mu^*$ was fitted to obtain $T_c$ at ambient pressure. Then for the variation with pressure they use the empirical relation of Bennemann and Garlandand\cite{Bennemann}. We got $\mu^*$ fitting it to $T_c$ through the LME equation at each pressure instead. So our values are consistent with the Mc Millan-Eliashberg linear equation and no further approximation is needed.   In a recent work, Profeta \textit{et all.}~\cite{Profeta} studied the behaviour of $T_c$ for Al as a function of pressure and obtained a good agreement with experiment. The experimental results we compare with are the ones  of Gubser and Webb \cite{Gubser} and Sundqvist and Rapp \cite{Sundqvist}. Our results are in excellent agreement with the ones of Profeta \textit{et all.}~\cite{Profeta} and in good agreement with experiment.  We present our results in Fig.~\ref{aluminio} and in the next Table I

\bigskip

\begin{table}[H]
\begin{center}
Table I - Properties of superconducting  Al under pressure
\end{center}
\begin{ruledtabular}
\begin{tabular}{c c c c c c c}

P[GPa]&	$T_c ^{FDM} [K]^a$ &	$T_c^{SCDFT} [K]^b$ & $T_c^{Exp}[K]^c$ &   $\lambda $ &  $\mu^* $   & a[Bohr] \\ \hline
0.0	&   $ 1.18 ^d $r & 1.18 & 1.18 & 0.4259  &  0.14154  &  7.6460\\
0.5	&    1.078 & 1.06 & 0.90 &  0.4168 &  0.14087  &  7.6297\\ 
1.0	&    0.969 & 0.96 & 0.79 &  0.4084  &  0.14024 &  7.6141\\ 
1.5	&    0.872 & 0.87 & 0.70  & 0.4009 &  0.13967  &  7.5989 \\
2.0	&    0.791 & 0.78 & 0.62  & 0.3942 & 0.13950   &  7.5841\\
2.5	&    0.706 & 0.71 & 0.56  & 0.3870  & 0.13914  &  7.5695\\
3.0	&    0.634 & 0.63 & 0.46  & 0.3805  &  0.13888  & 7.5553\\
3.5	&    0.567 & 0.57 & 0.37  & 0.3743  &  0.13872  & 7.5415\\
4.0	&    0.505 & 0.50 & 0.29  & 0.3680  & 0.13856  &  7.5280\\ 
4.5	&    0.448 & 0.42 & 0.22  & 0.3623  &  0.13876  & 7.5148\\
5.0	&    0.392 & 0.40 & 0.16  & 0.3569  &  0.13935  & 7.5019\\
5.5	&    0.324 & 0.38 & 0.10  & 0.3518  &  0.14190  & 7.4892\\
6.0	&    0.204 & 0.35 & 0.07  & 0.3460  &  0.15205  & 7.4766\\

\end{tabular}
\end{ruledtabular}
\caption{ a) our results b) {Profeta \textit{et all.}~\cite{Profeta}.} c) Lineal interpolation of the experimental data read from Gusber and Webb Ref.~\cite{Gubser}, and Levy and Olsen~\cite{Levy}. Similar figures can be found in  Sundqvist~\cite{Sundqvist} and Profeta \textit{et all.}~\cite{Profeta}. d) We took our input data from~\cite{Gubser}}.
\end{table}

In Table I, we consider a variation of pressure, $P$, from 0-6 GPa.  We first compare our results for $T_c$ as a function of pressure, $P$, with the ones  of  Profeta $\textit{et al.}$~\cite{Profeta}. The agreement is excellent. In the next column we present the result from experiment~\cite{Gubser,Levy}. The trent is reproduced quite well.  Next, we show the variation of the electron-phonon interaction parameter,  $ \lambda$. It always diminishes  with pressure. The decrement in the value of it is not exactly equal for all intervals of pressure since it varies from 0.0091 between $P=0$ and $P=0.5$ GPa to 0.0051 between $P=5$ to $ P=5.5$ GPa.   The electron-electron repulsion parameter, $\mu^*$, behaves somehow differently according to our calculations, since it presents a minimum. At 0 GPa, its value is 0.14154 and decreases steadily to a minimum value of 0.13856 at 4 GPa. Increasing the pressure,   $\mu^*$ increases and reaches a value of 0.15205 at 6 GPa. The minimum of the decrement in  $ \lambda$ arises between $P=5$ and $P=5.5$ GPa and so it does not correlate with the minimum in  $\mu^*$. The lattice parameter, $a$, diminishes steadily with pressure. Upon a 0.5 GPa enhancement in pressure it changes with a difference around 0.0148 Bohr. This decrement in the lattice constant is higher at low pressure and smaller at high pressure. The minimum occurs at $P=6$ GPa.  So, this behavior does not seem to correlate either with the behavior of the electron-electron repulsion parameter $\mu^*$. Further, if we look at the contribution of each phonon mode (two transverse and one longitudinal) to  the behavior of $T_c$ under pressure by taking only the corresponding energy interval of  $\alpha^2F(\omega)$ into account and apply to  this part only our method, we get the result  that they all contribute lowering the $T_c$. This behavior is not universal. Some preliminary results for Nb give evidence of a different behaviour.  

\section{Conclusions}
We presented in this paper an application of the functional derivative of the critical temperature with the Eliashberg function, ${\delta T_c}/{\delta \alpha^2F(\omega)}$, to calculate $T_c$ as a function of pressure. We applied the method to superconducting Al. We get an excelent agreement with the successful calculations of Profeta et al.~\cite{Profeta}  which are in agreement with experiment. This work can be extended to calculate the thermodynamics under pressure (the thermodynamic critical field, H(0), the jump in the specific heat and the gap, for example). This is the subject of our next work.

\section{Acknowlegments}
This  work  was  performed  using  the  facilities  of  the  super-computing center (Xiuhcoatl) at CINVESTAV-M\'exico.  Gonz\'alez-Pedreros acknowledges the support of Conacyt-M\'exico through a PhD scholarship.
\bibliography{PRBAl}
\end{document}